\documentclass[reprint,superscriptaddress,twocolumn,amsmath,amssymb,aps,prb]{revtex4}

\usepackage{graphicx}
\usepackage{amssymb}
\usepackage{amsmath}
\usepackage{nicefrac}

\begin{document}

\title{The contact mechanics challenge: Problem definition}



\author{Martin H. M\"user}


\affiliation{
John von Neumann Institut f\"ur Computing and 
J\"ulich Supercomputing Centre,
Institute for Advanced Simulation,
FZ J\"ulich, 52425 J\"ulich, Germany}
\affiliation{
Department of Materials Science and Engineering,
Saarland University, 66123 Saarbr\"ucken, Germany 
}

\author{Wolf B. Dapp}
\affiliation{
John von Neumann Institut f\"ur Computing and 
J\"ulich Supercomputing Centre,
Institute for Advanced Simulation,
FZ J\"ulich, 52425 J\"ulich, Germany}
\date{\today}

\begin{abstract} 
We present a contact mechanics problem, which we consider to be
representative for contacts between nominally flat surfaces. 
The main ingredients of the mathematically fully defined contact problem are:
Self-affine roughness, linear elasticity, the small-slope approximation,
and short-range adhesion between the frictionless surfaces.
Surface energies, elastic contact modulus and computer-generated
surface topographies are provided at www.lms.uni-saarland.de/contact-mechanics-challenge.
To minimize the undesirable but frequent problem of unit conversion errors, 
we provide some benchmark results, such as the
relative contact area as a function of load $a_{\rm r}(L)$ between 0.1\%
and 15\% relative contact. 
We call theorists and numericists alike to predict quantities that
contain more information than $a_{\rm r}(L)$ and provide information on how
to submit predictions. 
Examples for quantities of interest are the mean gap or contact stiffness as 
a function of load as well as distributions of contact patch size, interfacial 
stress, and interfacial separation at a reference load. 
Numerically accurate reference results 
will be disseminated in subsequent work 
including an evaluation of the submitted results. 
\end{abstract}

\keywords{Contact mechanics, Adhesion, Surface Roughness Analysis and Models}

\maketitle

\section{Introduction}

2016 marks the 50th anniversary of the pioneering approach by Greenwood
and Williamson (GW) to describe quantitatively the contact mechanics of 
nominally flat, but microscopically rough surfaces~\cite{Greenwood66}. 
The field still thrives, in part due to theoretical advances reducing a 
highly complex problem to one that can be handled on small-scale computers.
The arguably most prominent publications on contact mechanics since the GW paper 
are  the GW-inspired work of Bush, Gibson, and Thomas~\cite{Bush75} 
and the scaling theory proposed by Persson~\cite{Persson01}. 
There has also been much progress in the brute-force solution to contact 
mechanics.
It is now possible to simulate systems that are sufficiently large to mimic
the multi-scale nature of surfaces while reaching the continuum limit 
through sufficiently fine discretization at the small 
scale~\cite{Hyun04,Dapp14TL}. 

%

Comparisons between theoretical predictions and rigorous simulations 
--- making no uncontrolled approximations beyond the model assumptions ---
are usually limited 
to the question if a model reproduces the linearity between load
and contact area~\cite{Campana07,Carbone08,Ciavarella00,Hyun04,Paggi10}. 
Such comparisons are weak tests because theories merely need to 
reproduce a single proportionality coefficient while they usually depend on more 
than one adjustable parameter, which may not even be well defined from 
experiment or the model definition. 
The adjustable parameter thereby becomes effectively  a fit parameter. 
An important such term is the scale-dependent radius of curvature of an 
asperity~\cite{Greenwood01}, which plays a critical role in asperity-based
models. 

Comparisons of theories and rigorous simulations beyond the proportionality 
coefficient of load and true contact area have been scarce.
Notable examples are the analysis of the following quantities:
the gap distribution function~\cite{Almqvist11}, the dependence of mean gap 
or contact stiffness on load~\cite{Pastewka13PRE,Pohrt12}, or the 
interfacial stress spectrum~\cite{Campana08,Persson08JPCMb}.

So far, the in-depth comparisons between theory and rigorous
simulations have mainly focused on adhesionless contacts, while
adhesive interfaces have been even more scarce. 
%
The reason for this may be that modeling short-range adhesion in continuum
models places large demands on 
simulations while finite-range adhesion is usually more involved in 
theoretical approaches. 
In fact, handling short-range adhesion in simulations of single-asperity
contacts and reproducing (closely) the famous analytical results by Johnson, 
Kendall, and Roberts (JKR)~\cite{Johnson71} is not easy to achieve. 
A fine discretization is required close to the contact line~\cite{Muser14Beil}. 
Thus, a rigorous, numerical approach to short-range adhesion in mechanical
contacts, irrespective of it being based on a finite-element or boundary-value
method 
remains difficult.
%

In the following, we describe the model in section~\ref{sec:model}.
Selected results are presented in section~\ref{sec:results} while
section~\ref{sec:quantities} contains a list of quantities to be computed
and some discussion of these quantities. 
The final section~\ref{sec:conclusions} summarizes the rules for 
the competition.

\nopagebreak
\section{Model}
\label{sec:model}

\subsection{Surface Topography}

\begin{figure}[htb]
\includegraphics[width=0.45\textwidth]{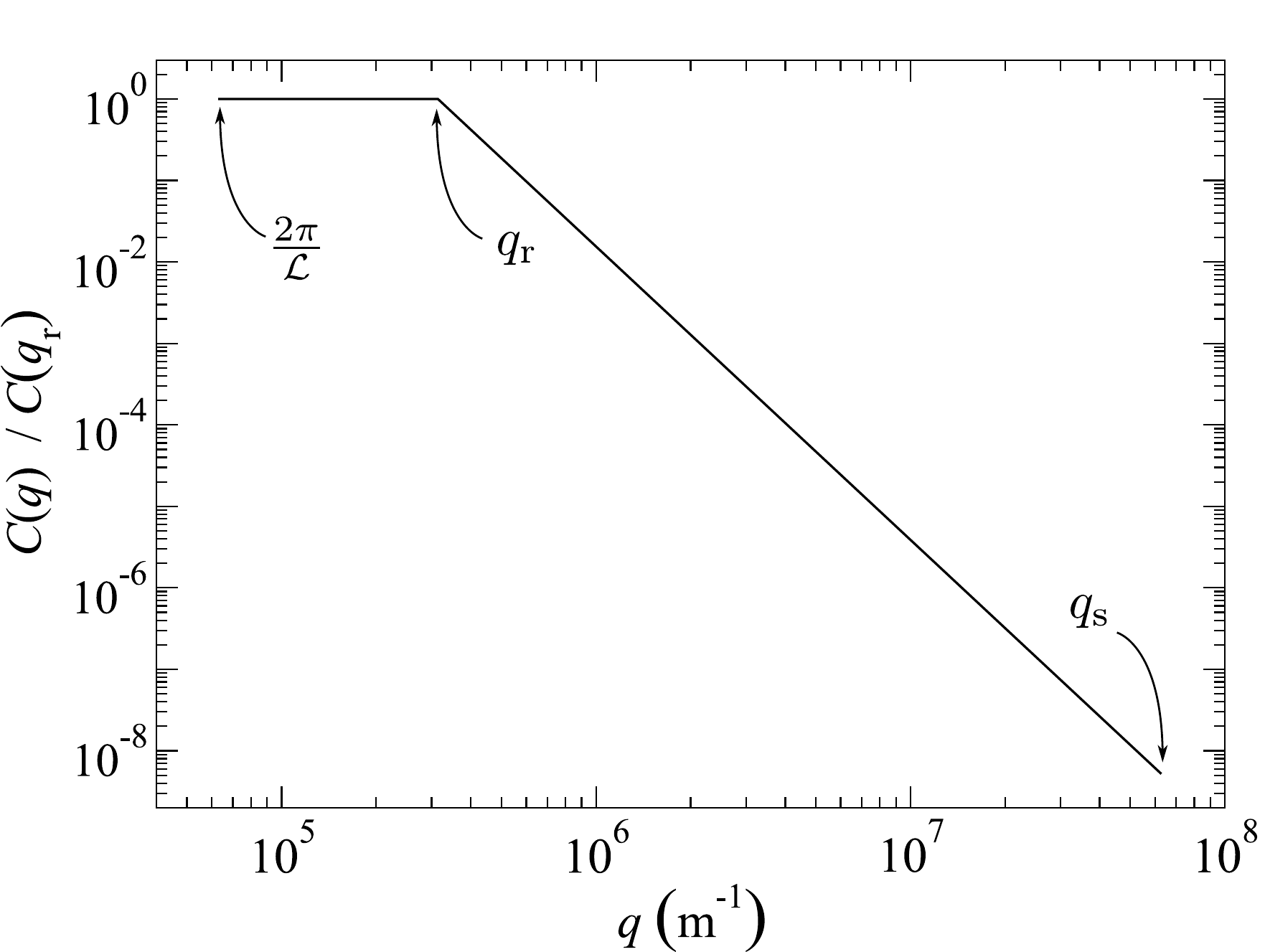}
\caption{
Height spectrum $C(q)$ from which the height distribution is drawn.
It is normalized to its value at the roll-off wavenumber $q_{\rm r}$.
\label{fig:spectrum}
}
\end{figure}

We produced our surface realization by drawing random numbers for the
Fourier transform of the height profiles $\tilde{h}({\bf q})$ having
a mean of zero and, on average, a second moment defined by the 
height spectrum

\begin{eqnarray}
C(q) & \equiv & \left\langle \vert \tilde{h}({\bf q}) \vert^2 \right\rangle \\
     & = & 
C(q_{\rm r}) \times
\left\{
\begin{array}{ll}
1 
  & \mbox{for } \lambda_{\rm r} < 2\pi/q \le {\cal L} \\
(q/q_{\rm r})^{-2(1+H)}
  & \mbox{for } \lambda_{\rm s} \le 2\pi/q < {\lambda_{\rm r}} \\
0           
  & \mbox{else.}
\end{array}
\right.\nonumber
\end{eqnarray}
Here, ${\cal L} = 0.1$~mm is the linear dimension in $x$ and $y$ of the 
periodically repeated simulation cell,
$\lambda_{\rm r} = 20$~$\mu$m is the roll-off wavelength, 
$q_{\rm r} = 2\pi/\lambda_{\rm r}$, and 
$\lambda_{\rm s} = 0.1$~$\mu$m is the short wavelength cutoff, below
which no roughness is considered.
Finally, $H$ is called the Hurst roughness exponent.
A graph showing the spectrum is presented in 
Fig.~\ref{fig:spectrum}. 
The features of the spectrum are similar to those summarized recently by 
Persson~\cite{Persson14TL}.

One might argue that introducing a cutoff at small wavelength is artificial.
However, we see this as necessary in order to make it possible to compare
simulations to continuum theories.
For similar reasons, we prefer a hard-wall interaction over finite-range repulsion.
Even if the latter might be more realistic and, in some ways, easier to handle
numerically (e.g., when relaxing the displacement field with a conjugate
gradient method), hard-wall repulsion allows us to define unambiguously 
points of contact and interfacial separation.

\begin{figure}[htb]
\includegraphics[width=0.45\textwidth]{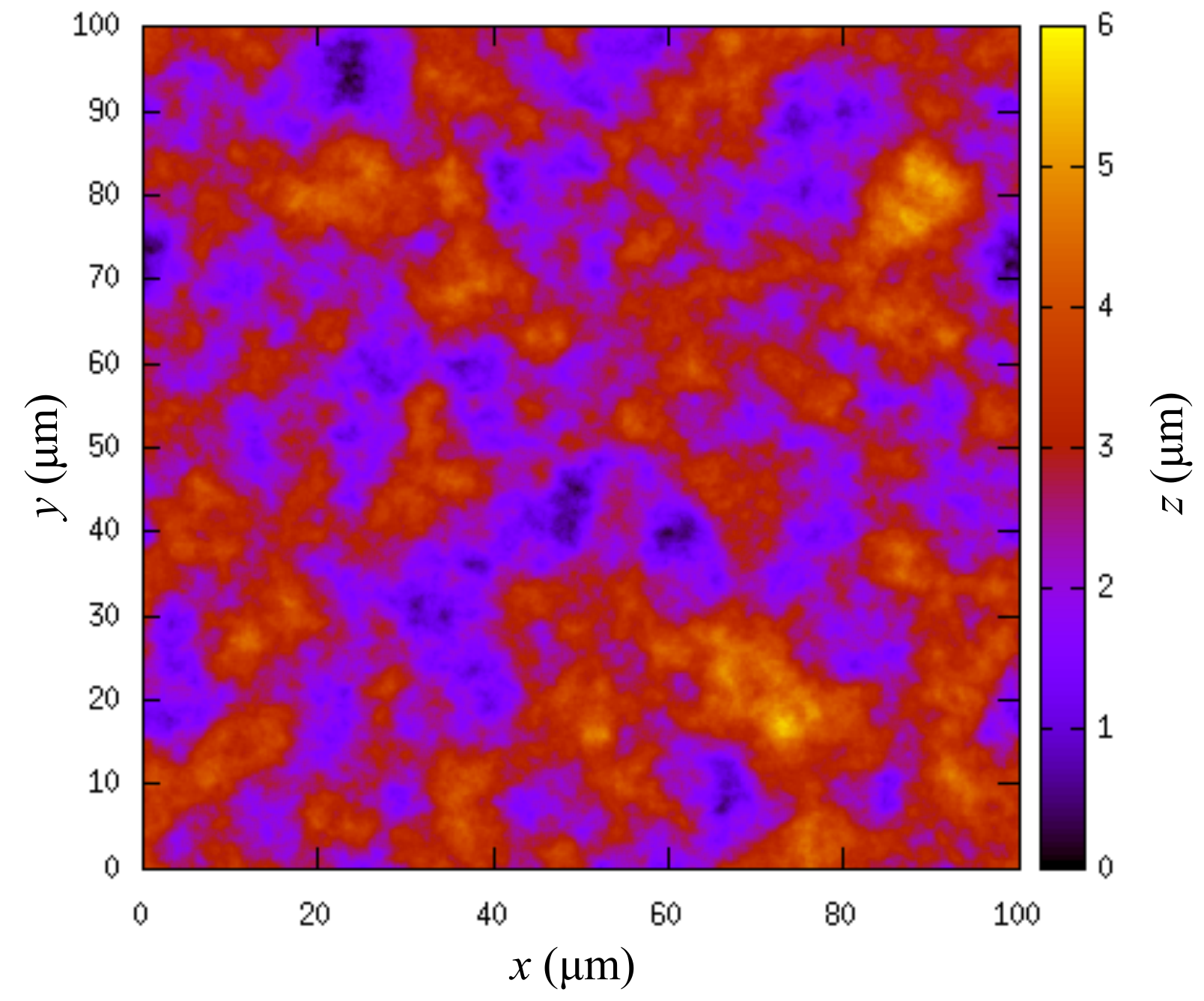}
\caption{
Height profile of the random surface that was produced from the
spectrum shown in Fig.~\ref{fig:spectrum}. 
\label{fig:surfaceTopo}
}
\end{figure}

The specific surface realization drawn from the spectrum is depicted in
Fig.~\ref{fig:surfaceTopo}.
We normalize the height spectra such that the root-mean-square gradient
of the height is $\bar{g}=1$.
Furthermore, we shift the heights such that the minimum value is zero.
Further characteristics of the surface topography are:
mean height $\langle h \rangle= 2.633$~$\mu$m, 
maximum height ${h}_{\rm max} = 5.642$~$\mu$m, and a 
root-mean square height fluctuation of 
$\sqrt{\langle \delta h^2 \rangle} = 0.762$~$\mu$m
with
$\langle \delta h^2 \rangle \equiv \langle h^2 \rangle - \langle h \rangle^2$. 
The inverse root-mean-square curvature, which one may interpret as a typical
local radius of curvature, is
$R_{\rm c} = 60$~nm. 

The surface is pressed down against an originally
flat elastic manifold. 
Thus, the first points of contact occur at small height,
i.e., in the dark areas of Fig.~\ref{fig:surfaceTopo}. 
The surfaces can be downloaded from the Internet.
Links are provided at 
http://www.lms.uni-saarland.de/contact-mechanics-challenge/. \\

\subsection{Elasticity, external load, and adhesion}

We assume the small-slope approximation, which forms the basis for essentially
any contact mechanics theory.
All roughness is mapped to the indenter, while all compliance is assigned to the 
substrate with a contact modulus of $E^* = 25$~MPa, which is characteristic for 
rubber.
Here, $E^* \equiv E/(1-\nu^2)$, where $E$ is the Young's modulus
and $\nu$ the Poisson ratio. 
We leave $\nu$ unspecified, as we focus exclusively on normal displacements. 

The external default pressure acts homogeneously across the system. 
It is set to 0.01~$E^*\bar{g}$=250~kPa.
In other words, the total load on the simulated area of 0.01~mm$^2$ is 0.25~N. 
The elastically deformable solid is assumed to be semi-infinite.
Mean displacements shall refer to that of the layer pressed against the 
counterface and not to the layer onto which one would exert external forces 
experimentally.

Short-range repulsion is realized with a hard-wall interaction, that is, 
the indenter is not allowed to penetrate the rigid substrate.
In addition, the two surfaces interact with a finite-range adhesion  
according to
\begin{equation}
v[g] = -\gamma_0 \int d^2r \exp\{-g({\bf r})/\rho\},
\end{equation}
where $\gamma_0 = 50$~mJ/m$^2$ is the surface energy at perfect contact,
$g({\bf r})$ is the local gap or interfacial separation as a function of the
in-plane coordinate ${\bf r}$, 
and $\rho = 2.071$~nm.
We note that the exponential cohesive zone model used here gives
essentially identical results to the analytical solutions of 
Maugis~\cite{Maugis92}, who used Dugdale's model for adhesion, see
Figs.~9 and 10 in Ref.~\onlinecite{Muser14Beil}. 

Defining a local Tabor coefficient according to
$\mu_{\rm T} \equiv R_{\rm c}^{1/3} (\gamma_0/E^*)^{2/3}/\rho$,
we obtain $\mu_{\rm T} = 3$. 
This value can certainly be classified as short-range adhesion.
This is evidenced in the two bottom panels of Fig.~\ref{fig:zoomIn}.
They reveal that the adhesive stress is rather localized near the 
contact lines. 
See also Figs.~9 and 10 in Ref.~\onlinecite{Muser14Beil} from where
it is also evident that $\mu_{\rm T}=3$ is close to the JKR limit of
infinitely short-ranged adhesion, at least as far as contact radius and
normal displacement are concerned. 

The parameters were chosen to mimic the contact between rubber and
a highly polished surface, though the contact modulus may be somewhat
at the upper end for practical applications. 
However, we set up the model such that there is no significant adhesive
hysteresis up to moderate contact pressures. 
Otherwise, functional relations such as $\bar{u}(L)$ or 
$a_{\rm r}(L)$ would become history dependent thereby impeding 
comparisons between theoretical predictions and our simulations. 

Pastewka and Robbins~\cite{Pastewka14} found that surfaces 
only became  hysteretic or ``sticky'' as long as the ratio of ``repulsive'' contact
area and load no longer increases linearly with pressure at small
contact area.
We found similar results~\cite{Muser16TI} and thus chose an adhesion such that 
the total contact
area is increased by roughly 50\% compared to the adhesionless case ---
at relative contact areas of a few percent.

\subsection{Summary of default parameters}
\label{sec:DefaultParams}

Two important dimensionless quantities of our default problem are the Tabor parameter 
$\mu_{\rm T} = 3$ and the surface root-mean square gradient $\bar{g} = 1$. 
Additional quantities in SI units are:
$E^* = 25$~MPa, $\gamma_0 = 50$~mJ/$m^2$, $\rho = 2.071$~nm,
system size ${\cal L} = 0.1$~mm, externally applied pressure $p_0 = 250$~kPa. 

It might be beneficial to use a problem-adopted unit system, 
which  is what we do in our own simulations.
In this unit system one has: $E^* \bar{g}$ as the unit for pressure
and ${\cal L}$ as the unit for length.
One can then use
$E^* = 1$,
${\cal L} = 1$, 
$p_0 = 0.01$,
$\gamma_0 = 2\times10^{-5}$,
and 
$\rho = 2.071\times10^{-5}$.

\subsection{Notes on our numerical solution}

We solve the contact model using the Green's function molecular dynamics (GFMD)
method as described in Ref.~\onlinecite{Dapp14TL}. 
The short-ranged adhesion puts large demands on the discretization.
Reaching convergence necessitates fine discretization, in particular for
adhesive necks forming near contact lines.
We found that a discretization of $a = O(\lambda_{\rm s}/64)$ is sufficient 
for most purposes and consequently produce reference data on systems with
$2^{16}\times 2^{16} \approx 4\times 10^{9}$
discretization points on the surface. 
In some cases, we use $a = O(\lambda_{\rm s}/128)$ or $\approx 16\times 10^{9}$ 
grid points to ensure that results have sufficiently closely converged to the
continuum limit. 
Using a fine-tuned value for the damping, 
the system can be typically relaxed within a few thousand time steps,
although equilibration at the smallest investigated loads, leading to
0.3\% relative contact, necessitates roughly ten times longer simulations. 
%

\section{Selected Results} 
\label{sec:results}

\subsection{Configurations at the reference load}

We present some selected results, in order to give theorists and modelers the 
opportunity to ensure that results are in the correct ballpark and/or that 
units have been implemented correctly. 
However, it is of no interest to us to receive explanations for why a given
theory may be consistent with the provided results.
Only true predictions, rather than ``post''-dictions, can enter the competition 
and only predictions will be credited in a future paper assessing the value 
of the submitted data.

\begin{figure}[htb]
\begin{center}
\includegraphics[width=0.475\textwidth]{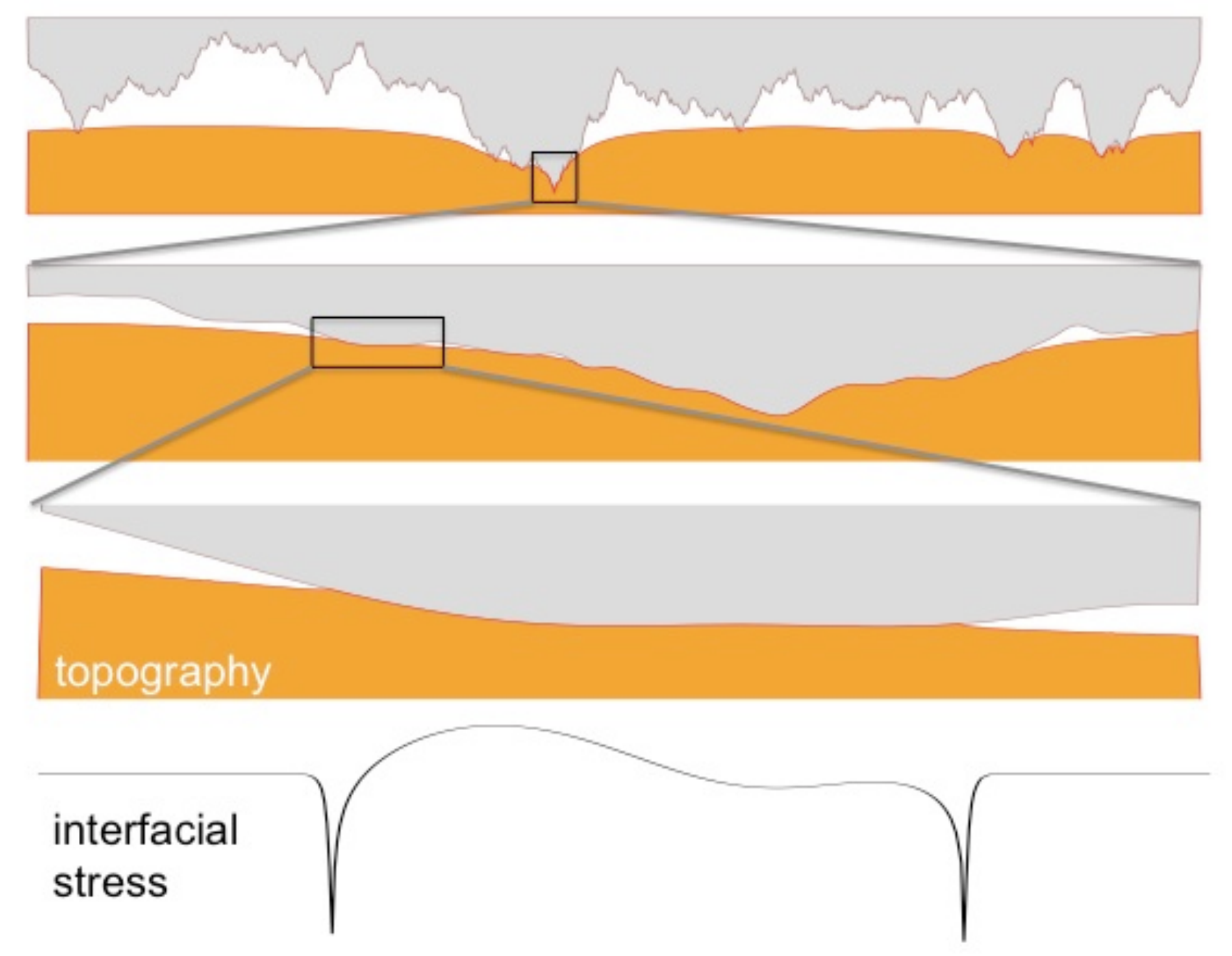}
\end{center}
\caption{
\label{fig:zoomIn}
Contact geometry at different magnifications (three top panels) and 
local interfacial stress (bottom panel) for the reference geometry.
The cut runs through the system shown in Fig.~\ref{fig:surfaceTopo} at
$x = 50$~$\mu$m. 
Normal and lateral scale differ in all panels. 
Rectangles highlight the region shown at larger magnification one panel below. 
Values for adhesion and load differ from their default values by some factor.
%
Only the Tabor coefficient was kept at its reference value of
$\mu_{\rm T} = 3$. 
}
\end{figure}

We first show the gap geometry as well as the interfacial stress in a cross section
through our system in Fig.~\ref{fig:zoomIn}. 
We took the geometry as defined above but multiplied the default values for adhesion 
and load by some factor to make it difficult for participants of the competition 
to deduce results for the gap.
Figure~\ref{fig:zoomIn} should nevertheless give an impression of the 
problem's complexity.
In particular, it shows the need of a fine discretization near contact lines. 
In this model, material points at the contact line experience the highest 
tensile stresses of magnitude $\gamma_0/\rho$.

\subsection{Relative contact area as a function of load}

\begin{figure}[htb]
\begin{center}
\includegraphics[width=0.475\textwidth]{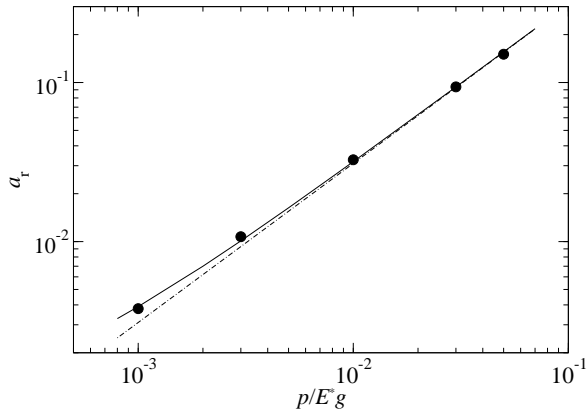}
\end{center}
\caption{
Relative contact area $a_{\rm r}$ as a function of the reduced
pressure $\tilde{p} = p/E^*\bar{g}$ for the default system. 
The dashed line shows a linear relation, while the solid line is meant
to guide the eye.
\label{fig:heating1}
}
\end{figure}

The relative contact for moderate loads is shown in 
Fig.~\ref{fig:heating1}.
The default load or pressure corresponds to the value
$\tilde{p} \equiv p/E^*\bar{g} = 0.01$.
The proportionality coefficient between $a_{\rm r}$ and $\tilde{p}$
is slightly greater than three, i.e., it is increased by roughly
50\% as compared to the adhesionless case.

\vspace*{-4mm}
\vspace*{-4mm}

\section{Quantities to be predicted}
\label{sec:quantities}

In this competition, we focus on physical observables that derive
from the interfacial separations field and the stress field. 
Since we can only run a limited number of simulations for comparison,
we need to restrict the number of pertinent observables. 
We therefore chose one reference point, in which all variables describing
the system are fixed and ask competitors to predict various properties
that can be defined on that system.
We also allow for predictions on how selected observables {\bf vary}
with {\bf one single parameter keeping all others fixed}
at their values defined in section~\ref{sec:DefaultParams}.

\begin{enumerate}
\item Distribution of contact patch size.\\
Suitable representations for this observable are used in Figs.~4 
and~5 of Ref.~\onlinecite{Campana08alone}. 
\item Contact area as a function of load or pressure.\\
Here, we are interested in data for $a_{\rm r}$ approaching unity,
i.e., above 75\% contact. 
Moderate contact areas can too easily deduced by extrapolating 
results presented in this work. 
To reveal the asymptotic behavior at large loads, a representation
of the $a_{\rm r}(p)$ relationship as shown in Fig.~1 of 
Ref.~\onlinecite{Dapp14b} is certainly acceptable. 
\item Gap distribution function.\\
As an example we refer to Fig.~9  of Ref.~\onlinecite{Almqvist11}. 
Please use logarithmic axes when asymptotic behavior might be unclear
otherwise. 
\item Interfacial stress distribution function.\\
Here, we ask to plot the stress on a linear scale and the probability to 
find a given stress on a logarithmic scale similar to the way how it is
done in Fig.~9 of Ref.~\onlinecite{carbone09}. 
\item Interfacial stress spectrum.\\
Again, we can only consider calculations for the default parameters. 
Examples are given in Figs.~6--8 of Ref.~\onlinecite{Campana08}. 
\item Contact area as a function of the Tabor coefficient.\\
All model parameters other than $\mu_{\rm T}$ shall be kept constant.
The submitted results should be similar in representation as Fig.~9
in Ref.~\onlinecite{Muser14Beil}, except, of course, that the contact radius
has to be replaced with (relative) contact area and the single-asperity
geometry with the given geometry.
Please use a logarithmic axis for $\mu_{\rm T}$. 
The interesting behavior is found for Tabor parameters less than the 
default value, i.e., when short-range adhesion crosses over to long-range
adhesion. 
As a clue to the solution, we reveal that the behavior is monotonic
but ``interesting'' around the point where the mean gap has a local
extremum. 
\item Mean gap or normal displacement (of the layer pressed against the
counterface)  as a function of the Tabor coefficient. \\
Again, all quantities but the Tabor coefficient (or interaction range)
should be kept at their default values.
Here we refer to Fig.~10 in Ref.~\onlinecite{Muser14Beil} and ask again
to use a logarithmic representation for the Tabor coefficient. \\
A brief discussion might be helpful: As in single-asperity contacts, 
the mean gap (displacement) initially decreases (increases in magnitude)
when $\mu_{\rm T}$ is reduced from its reference value. 
However, for the investigated contact geometries, this trend
must eventually reverse, as for finite or periodically repeated
contacts, the adhesive extra load (per unit area) disappears when
$\mu_{\rm T}$ approaches zero.
Finding the value of $\mu_{\rm T}$ at which the mean gap has its local 
extremum is a ``challenge in the challenge'', even if the adhesive interaction 
range at which this extremum is located, 
might not be practically relevant. 
\item Mean gap or normal displacement as a function of load.\\
This quantity has been predicted numerous times by various authors for
non-adhesive surfaces, for example, in Fig.~5 of Ref.~\onlinecite{Almqvist11}.
In that plot it would have been beneficial to use a logarithmic representation
for the abscissa, as it would have allowed one to see the asymptotic behavior
at small mean separation. \\
We do not ask for the interfacial stiffness as a function of load as it
follows from the mean gap by differentiation. 
Thereby, it does not provide new information. 
\nopagebreak
\item Gap and stress along a cross section. \\
Participants using numerical methods, e.g., homogenization, are invited
to submit results on gap and stress (with units) for the cross section at 
$x = 50$~$\mu$m shown in Fig.~\ref{fig:zoomIn}. 
We can smooth the true (microscopic) stress to make a meaningful comparison
to homogenization or related approaches. 
\end{enumerate} 


\section{How to submit}
\label{sec:conclusions}

Please summarize your results in a PDF, which you e-mail to:
martin.mueser@mx.uni-saarland.de.
PDFs should not exceed 5~MB and be sent by May 31, 2016.
The number of figures is limited to ten.
Their format should be similar to the ones referred to in the previous
section.
You are also invited to describe the method with which your data was
produced.
It will help us to advertise your results and your method if the description 
is to the point, i.e., complete but brief. 

If your simulations are deemed successful (committee: Prof. Martin M\"user, 
Prof. Jim Greenwood, Prof. Wilfred Tysoe, and Prof. Nicholas Spencer), 
we offer to present your data in a publication in Tribology Letters. 
If you agree to having your data disseminated and if you agree to our write-up, 
we will ask for your data (which we receive preferentially as ascii).
Once your data has been received and processed, 
you will be offered co-authorship. 
If you disagree, we will not use your data.

\bibliographystyle{unsrt}
\bibliography{seal12}

\end{document}